# Control Strategy for Anaesthetic Drug Dosage with Interaction Among Human Physiological Organs Using Optimal Fractional Order PID Controller

Saptarshi Das[1,2]
1. School of Electronics and Computer Science, University of Southampton, Southampton SO17 1BJ, UK.
Email: s.das@soton.ac.uk, saptarshi@pe.jusl.ac.in

Sourav Das[2] and Koushik Maharatna[1]
2. Department of Power Engineering, Jadavpur University, LB-8, Sector 3, Kolkata-700098, India.
Email: das_sourav@live.in, km3@ecs.soton.ac.uk

*Abstract*—In this paper, an efficient control strategy for physiological interaction based anaesthetic drug infusion model is explored using the fractional order (FO) proportional integral derivative (PID) controllers. The dynamic model is composed of several human organs by considering the brain response to the anaesthetic drug as output and the drug infusion rate as the control input. Particle Swarm Optimisation (PSO) is employed to obtain the optimal set of parameters for PID/FOPID controller structures. With the proposed FOPID control scheme much less amount of drug-infusion system can be designed to attain a specific anaesthetic target and also shows high robustness for ±50% parametric uncertainty in the patient's brain model.

## I. INTRODUCTION

Control strategy formulation for anaesthetic drug dosage is very crucial in clinical surgery and falls in a particular category of biomedical system design known as pharmacology [1-3]. In pharmacology, there are two major steps involved known as pharmacokinetics and pharmacodynamics [4]. The anaesthetic drug injected in a patient's body gets infused in the arterial blood flow and then the arterial blood carrying the drug reaches to different physiological organs. Determining the drug concentration in arterial blood flow and in different tissues i.e. dosage to concentration is known as the pharmacokinetics. The interaction of the drug with different physiological organs and the overall effect of the drug i.e. concentration to effect is known as pharmacodynamics [4]. Amongst many others anaesthetic drugs, fentanyl is widely used in relief of acute pain like cancer [5] and in different surgeries [6]. The drug dosage in clinical surgery is generally controlled by the Electroencephalogram (EEG) recording during the anaesthesia [7] until a predefined unconsciousness in not reached which can be characterized by observing slow oscillations in EEG signals. Since different physiological organs have different time constants to absorb, react and finally release the drug again in the blood stream, their interaction and contribution on the overall physiological dynamics of anaesthetic study is highly complex and may lead clinicians to misinterpret the observed event. As an example, some organs may store the drug in a larger extent or reacts to the drug slowly than the others and its immediate effect on the reduction of fast oscillations in EEG waves may not be significant. This type of phenomena may confuse the clinicians to enhance the drug dosage to attend a pre-specified EEG activity indicative to an anaesthetic state in a typical case which might be life-threatening for the patient. In this scenario a physiological model based simulation study is necessary to device a control strategy with possible variation in patient's mathematical model from the nominal case, since over dosage of the drug by an open-loop type EEG observation based control may endanger the patient in the process of anaesthesia. A realistic (data-based) model for control strategy formulation can thus be implemented for automated anaesthesia in clinical surgery using brain activity (EEG) monitoring as a sensory feedback to the comparator to generate the tracking error. This results in a control action going to an actuator to pump the drug into patient's body [8], using a prior knowledge of the set-point or reference as quantitative measure of consciousness [4, 8].

Therefore the task of the controller design in the present scenario can be summarised as minimising the tracking error of the brain response given by the Hill equation, modified Hill equation (in frequency domain) and bispectral index (BIS) of EEG by delicately manipulating the drug input to the patient [4]. A realistic mathematical model for fentanyl drug was developed in [9-10] and is known as Mapleson-Higgins model. The Mapleson model was composed of a few set of algebraic equations derived from biochemistry. In this model, each organ of human body like lungs, peripheral shunt, kidney, gut and spleen, liver, other viscera, muscle, fat, sample brain is considered as a separate compartment. It is also assumed that each of the organs gets equal arterial blood flow. Mahfouf *et al.* [11] translated the static algebraic equation based Mapleson model into a dynamic model which considers the temporal variation in the drug concentration at the inlet and outlet of each physiological organ. They also carried out model reduction of the large dynamical system for generalized predictive control (GPC) design. In this paper, we use the original (unreduced) higher order dynamic model in order to design efficient control scheme. It has been shown by Das *et al.* [12]-[13] that fractional order (FO) controllers are very effective in handling higher order dynamics due to their inherent infinite-dimensional nature over integer order controllers which formulates the scope of the present study. Control strategies for anaesthetic dosage have been formulated on much simpler models using PID controllers [14]-[16], fuzzy

[17] and neuro-fuzzy [18] controllers in earlier research. Wada et al. [19] developed a more detailed physiological system level pharmacokinetic model without any control scheme. A three state nonlinear compartmental model for clinical pharmacology has been described for different control studies including adaptive control [20], neural network control [21-22], nonnegative dynamical systems [23], disturbance rejection control [24]. Results of clinical trials of anaesthesia control scheme, based on three-compartmental model with noise EEG measurements has been reported in Haddad et al. [25] for 10 patients. The focus of the present study is to formulate an optimal FO control strategy [26-28] with the dynamical model reported in Mahfouf et al. [11] considering interaction amongst physiological organs.

The rest of the paper is organised as follow: section II describes the overall system model for different physiological organs and their interaction with the drug. The optimal controller design task is described in section III and the system simulation in section IV. The paper ends with the conclusion in section V, followed by the references.

## II. DYNAMIC MODEL OF FENTANYL INTERACTION WITH PHYSIOLOGICAL ORGANS

### A. Overall System Description for Automated Anaesthesia

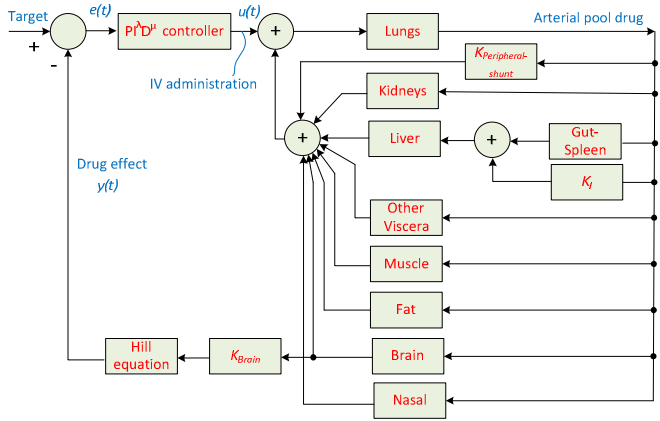

Fig. 1. Overall fentanyl dose-effect model with proposed control strategy.

As described earlier, the first static model of fentanyl interaction from pharmacokinetics and pharmacodynamics point of view was known as Mapleson model [9]-[10]. Mahfouf et al. [11] extended the concept for dynamic models by representing each organs dynamics with ordinary differential equations or single-input-single-output (SISO) continuous time transfer function models. The model mainly captures the drug flow from one organ to other and not the drug concentration in a specific organ. The drug is added through intravenous (IV) injection which gets infused in the arterial blood and then is perfused in all organs with a fraction of the total arterial blood flow. The brain reaction to the drug $y(t)$, manifested in the form of EEG is evaluated using the Hill-equation or some index like BIS and then fed-back for comparison with the target anaesthetic level to generate an error $e(t)$ which will be minimised by the FOPID or $PI^\lambda D^\mu$ controller to generate a control action $u(t)$, giving a command to the actuator or mechanical pump to pump the drug (Fig. 1).

### B. Dynamical Model for Each Human Organs

Mahfouf et al. [11] developed individual system models by applying system identification techniques on original Mapleson model with the consideration of 70 kg body weight and 6.48 l/min cardiac output which represent a base-case clinical scenario. In order to study the generalising capability of the model outside the nominal range, an interpolation based model identification has been reported in [11] for realistic variation either in body-weight, cardiac output or simultaneously both of them. From the generalized interpolated scheme, a patient model was developed with 93 kg body weight, 5.41/min body weight while 100μg of fentanyl has been injected over a period of 60 sec. The corresponding interpolated transfer function models [11] of various human organs are described in (1)-(9).

$$G_{\text{Fat}} = \frac{-1.437 \times 10^{-8} s + 1.722 \times 10^{-6}}{s^2 + 0.4126 s + 0.0003241} \quad (1)$$

$$G_{\text{Lungs}} = 64.78 / (s^2 + 4.17 s + 6.97) \quad (2)$$

$$G_{\text{Gut-spleen}} = \frac{-1.34 \times 10^{-5} s + 0.001604}{s^2 + 0.9059 s + 0.09356} \quad (3)$$

$$G_{\text{Kidneys}} = 0.0132 / (s + 0.7436) \quad (4)$$

$$G_{\text{Liver}} = 0.006243 / (s + 0.04257) \quad (5)$$

$$G_{\text{Other-viscera}} = 0.001725 / (s + 0.09009) \quad (6)$$

$$G_{\text{Muscle}} = \frac{-2.764 \times 10^{-7} s + 3.312 \times 10^{-5}}{s^2 + 0.4152 s + 0.001867} \quad (7)$$

$$G_{\text{Brain}} = 1.614 \times 10^{-5} / (s + 0.1533) \quad (8)$$

$$G_{\text{Nasal}} = 5.459 \times 10^{-6} / (s + 0.08507) \quad (9)$$

The peripheral shunt has been modelled as only as a gain without any time constants i.e. $K_{\text{peripheral-shunt}} = 0.0241$. The above models explain the dynamic relationship between the outgoing drug concentration in the blood flow from each tissue and the incoming arterial pool drug amount whereas it cannot model the drug concentration in each issue. If the drug concentration in the outgoing flow from brain is represented as $C_b$, then the drug effect on the brain be calculated by the Hill equation (10) using the effect-site concentration at 50% of drug effect ($EC_{50} = 7.8$ ng/ml) and steepness or slope factor ($\gamma$).

$$\text{Effect}(C_b) = C_b^\gamma / (EC_{50}^\gamma + C_b^\gamma), \quad \gamma = 4.3 \quad (10)$$

## III. OPTIMAL CONTROL SCHEME OF DRUG INFUSION WITH FRACTIONAL ORDER PID CONTROLLER

### A. Choice of the Control Objective

In clinical practices of anaesthesia, computer controlled drug infusion is generally adopted to meet a target concentration infusion (TCI) rather than manually controlling the infusion rate [8, 14, 15, 22, 25]. Most of these techniques rely on open loop control which assumes that the population model is a good representative of any patient which may not be valid in many cases. A more sensible scheme should be to continuously monitor brain response (EEG) to the drug-dosage and use it as a feedback mechanism to minimise the set-point tracking error by an efficient controller structure. The control signal essentially modifies the intravenous drug infusion rate and should not be violently manipulated to reach a faster anaesthetic effect. So apart from minimizing the tracking error $e(t)$, it is also an important task to minimize the variation (or derivative) of the drug infusion rate to prevent any sudden shock in the automated IV injection pump and chance of infusing large amount of drug in small time i.e. an increased control effort $u(t)$. The objective function for the optimal controller design has been formulated as (11) as a weighted sum of Integral of Time multiplied Squared Error (ITSE) and Integral of Squared Deviation Control Output (ISDCO).

$$J = \int_0^\infty \left( w_1 \cdot t \cdot e^2(t) + w_2 \left( \Delta u(t) \right)^2 \right) dt \quad (11)$$

*B. Controller Structure and its Tuning Using PSO Optimiser*

Fractional order $PI^\lambda D^\mu$ controller design with various other time domain performance criteria has been explored in Das *et al.* [12] and the results shows that squared error term in (11) puts more penalties on the tracking error and the time multiplication term makes the overall response faster and reduces the chance of loop oscillations in later stages. It is evident that such a tracking criterion will definitely increase the required control effort whose variation (temporal derivative) is thus added as a minimising criteria in (11). The above control objective is to be met by an integer and fractional order PID controller structure (12) where the controller gains $\{K_p, K_i, K_d\}$ and the integro-differential orders $\{\lambda, \mu\}$ are to be tuned with a global optimisation algorithm.

$$C(s) = K_p + \left( K_i / s^\lambda \right) + K_d s^\mu \quad (12)$$

For PID controller only the gains are to be optimized by considering the orders as unity. In (12), each FO operator is continuously rationalised within the optimization process using a 5$^{th}$ order Oustaloup's recursive approximation or ORA (13) for a chosen frequency band of $\omega \in \{\omega_l, \omega_h\} = \{10^{-2}, 10^2\}$ rad/sec [29] and the rationalised pole-zero and gains are given by $\{K, \omega_k, \omega'_k\}$.

$$s^\gamma \simeq K \prod_{k=-N}^{N} (s + \omega'_k) / (s + \omega_k)$$
$$\omega_k = \omega_b \left( \omega_h / \omega_b \right)^{\frac{k+N+(1+\gamma)/2}{2N+1}}, \omega'_k = \omega_b \left( \omega_h / \omega_b \right)^{\frac{k+N+(1-\gamma)/2}{2N+1}}, K = \omega_h^\gamma \quad (13)$$

Here, the PID/FOPID controller parameters are optimized using the PSO algorithm which is a widely used global optimiser. The swarm starts with particles having velocity $v_i$ and position $x_i$ and in each time step they try to move towards the global best with latest value of the best found solution for individual particle or *pbest* ($p_i$) and that of the global swarm or *gbest* ($p_g$), while the velocity and positions are manipulated over successive iterations using (14), until all particles converges to the global best solution.

$$v_i(t+1) = \omega v_i(t) + c_1 \varphi_1 \left( p_i(t) - x_i(t) \right) + c_2 \varphi_2 \left( p_g(t) - x_i(t) \right)$$
$$x_i(t+1) = x_i(t) + v_i(t+1) \quad (14)$$

Here, $\{c_1, c_2\} = \{0.5, 1\}$ are known as the inertia factor, cognitive learning rate and social learning rate and $\{\varphi_1, \varphi_2\}$ are uniformly distributed random variables within the interval $[0,1]$. The parameter $\omega$, known as the inertia factor for the swarm is linearly varied from 0.9 to 0.1. In the present study, the unconstrained version of PSO is employed with only bound on the controller parameters as $\{K_p, K_i, K_d\} \in [0.001, 10]$ and $\{\lambda, \mu\} \in [0, 2]$. Due the implementation issues of ORA for FO operators for $\{\lambda, \mu\} > 1$, we explored four classes of FOPID structures (15) and tested the tracking performance and control effort of each controller structure.

$$\text{FOPID}_1 \rightarrow (\{\lambda, \mu\} < 1); \quad \text{FOPID}_2 \rightarrow (\lambda < 1, \mu > 1);$$
$$\text{FOPID}_3 \rightarrow (\lambda > 1, \mu < 1); \quad \text{FOPID}_4 \rightarrow (\{\lambda, \mu\} > 1) \quad (15)$$

IV. SIMULATION, RESULTS AND DISCUSSION

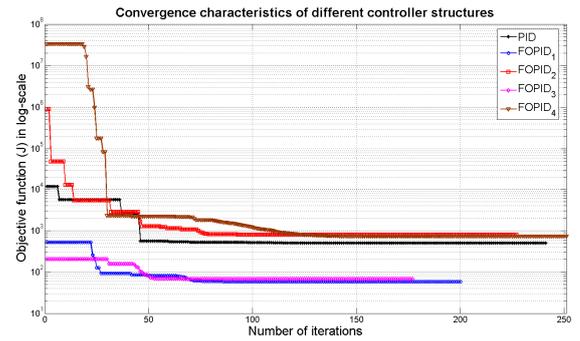

Fig. 2. Convergence characterstics of different controller structures.

TABLE I. OPTIMUM VALUES OF THE CONTROLLER PARAMETERS

| Controller | $J_{min}$ | Controller parameters | | | | |
|---|---|---|---|---|---|---|
| | | $K_p$ | $K_i$ | $K_d$ | $\lambda$ | $\mu$ |
| PID | 499.2007 | 3.8243 | 8.6647 | 0.001 | - | - |
| FOPID$_1$ | 58.3018 | 0.0212 | 2.3014 | 0.0783 | 0.8301 | 0.1013 |
| FOPID$_2$ | 792.1196 | 5.5757 | 3.3381 | 0.001 | 0.798 | 0.1488 |
| FOPID$_3$ | 68.721 | 0.2106 | 1.5393 | 0.001 | 0.001 | 0.021 |
| FOPID$_4$ | 728.6634 | 0.4213 | 1.6329 | 0.001 | 0.5237 | 0.1033 |

The PSO based optimal parameter selection is carried out for the 5 controller structure with a goal of minimizing the objective function (11) by considering a step input target at

$t = 1$ min and has been reported in Table I. The located minima ($J_{min}$) in Table I show that the FOPID$_1$ structure with $\{\lambda, \mu\} < 1$ gives the most optimal result than other FOPID variants and PID controller which is counter-proved by the PSO convergence characteristic, corresponding to FOPID$_1$. In most of the case low derivative gain implies that a PI/PI$^\lambda$ type controller is more suitable for this particular system.

### A. Performance on Nominal Patient Model

With the optimal parameters of the PID/FOPID controller in Table 1, the nominal anaesthetic drug delivery system, described in (1)-(10) and Fig. 1 is now simulated and the tracking performance and required control efforts are compared in Fig. 3 and Fig. 4 respectively. It is evident that although the with PID controller the tracking is much faster, it needs more amount of drug infusion within a short period of time (5 mins). Whereas the FOPID$_1$ structure meets the same anaesthetic target while pushing much less amount of drug into patient's body over a longer period of time (20 mins).

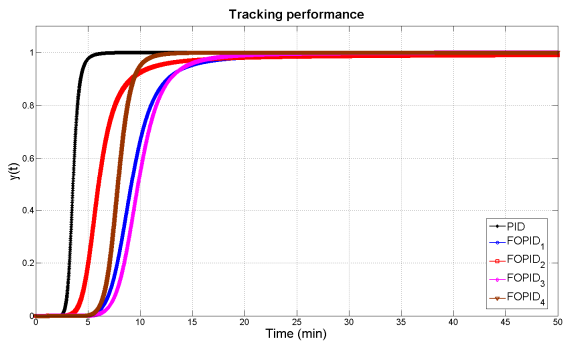

Fig. 3. Tracking performance for different controller structures.

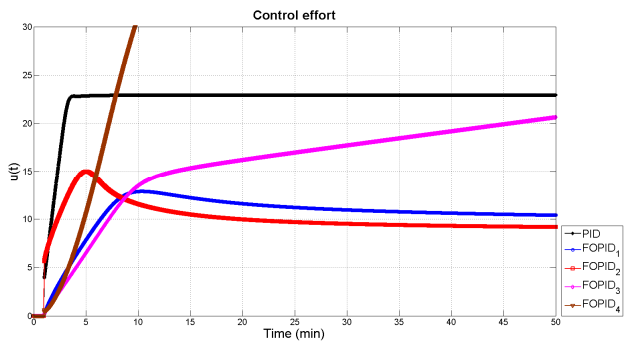

Fig. 4. Control effort for different controller structures.

### B. Robustness of Control Scheme in Perturbed Condition

In the previous subsection, the simulations have been reported for the nominal model whereas it is highly likely that the model differs significantly for different patients and for different conditions of the same patient. Since the brain model directly affects the output of the system as shown in Fig. 1, i.e. the brain response to fentanyl drug directly feedback to the controller, we here report simulation studies of the brain response for the drug in a ±50% dc gain perturbation scenario.

Testing vulnerability of control loops with dc gain variation is a widely used robustness measure and has been described in Das et al. [12]. The dc gain of the nominal brain model is $1.0528 \times 10^{-4}$. The rest of the physiological organs i.e. peripheral-shunt, kidney, liver, other-viscera, muscle, fat, nasal receive the same arterial blood flow in Mapleson model described in Fig. 1, but does not directly affect the output $y(t)$. Hence, they will have much less influence if the respective parameters are perturbed. Fig. 5-6 reports simulations with the best found cases of PID and FOPID controller respectively under a ±50% dc gain perturbation in the brain model (8).

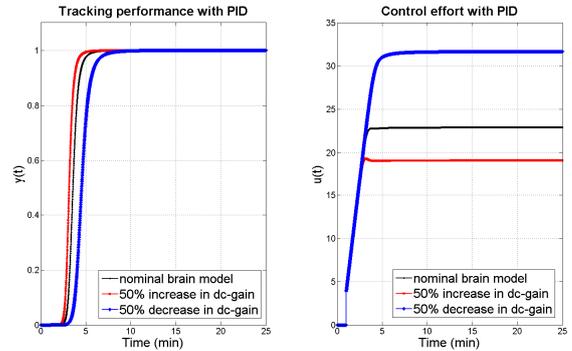

Fig. 5. Tracking performance and control effort with PID controller for ±50% dc-gain perturbation of the brain model.

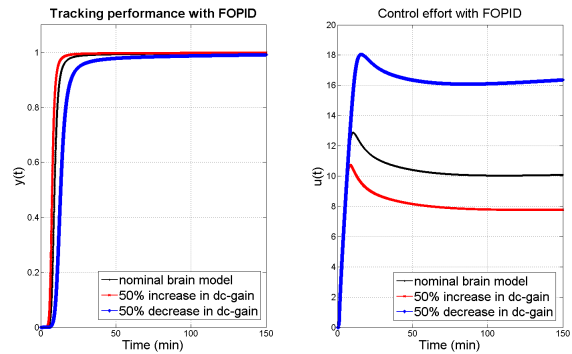

Fig. 6. Tracking performance and control effort with FOPID controller for ±50% dc-gain perturbation of the brain model.

It is evident from Fig. 5 that for a decrease in the dc-gain of nominal brain model, the control effort rapidly increases, signifying that more drug is injected into the patient's body. It is evident that the rise time of the PID controlled system is much faster (10 mins), but the amount of drug-injected is significantly high (17-35 units). On contrary, the FOPID controller in Fig. 6 provides a relatively slower tracking performance (50 mins). But there is a significant amount of saving in the drug injection level which is bounded within 7-18 units even in the perturbed condition. In clinical practices, meeting the specified unconsciousness level is not the sole criterion because this may create a sudden shock in different human physiological systems which may be harmful for the patient [8]. Also, in order to attain same level of anaesthetic effect but within a shorter time-period, the drug concentration in brain needs to be raised rapidly which has both economical constraint and physiological ill-effects along with faster mechanical pumping needed for automated drug-delivery. Whereas a smoother and bounded control action with less amount of drug infusion into the patient can attain the same level of anaesthethetic fentanyl drug, if equipped with an efficient control strategy i.e. using proposed optimal FOPID controller.

V. CONCLUSION

The paper devises a new fractional order control strategy for automatic fentanyl drug dosage control for anaesthesia in clinical practices. The proposed FOPID control scheme requires less amount of dug to be injected than with a PID controller to meet same anaesthetic target. The smoother control action provided by FOPID controller outperforms classical PID controller in a sense of restricting the chance of feeling a sudden shock in the brain response due to rapid increase in the amount of anaesthetic drug concentration in arterial blood flow within a very short time interval. Future work may be directed towards validation of the control scheme applied on Mapleson's dynamic model, with real clinical data.